\documentclass[prd,twocolumn,showpacs,preprintnumbers,amsmath,amssymb,superscriptaddress,floatfix,nofootinbib]{revtex4}

\usepackage{graphicx}
\usepackage{bm}
\usepackage{amsmath}
\usepackage{amsfonts}
\usepackage{amssymb}
\usepackage{color}
\usepackage{multirow}
\usepackage[colorlinks, citecolor=blue,anchorcolor=red,menucolor=red, linkcolor=red,filecolor=red,runcolor=red,urlcolor=blue,frenchlinks=red]{hyperref}

\begin{document}
\title{Assignments of the $Y(2040)$, $\rho(1900)$, and $\rho(2150)$ in the quark model }

\author{Zheng-Ya Li}
\affiliation{School of Physics and Microelectronics, Zhengzhou University, Zhengzhou, Henan 450001, China}

\author{De-Min Li}
\email{lidm@zzu.edu.cn}
\affiliation{School of Physics and Microelectronics, Zhengzhou University,
             Zhengzhou, Henan 450001, China}\vspace{0.5cm}

\author{En Wang}
\email{wangen@zzu.edu.cn}
\affiliation{School of Physics and Microelectronics, Zhengzhou University,
             Zhengzhou, Henan 450001, China}\vspace{0.5cm}

\author{Wen-Cheng Yan}
\email{yanwc@zzu.edu.cn}
\affiliation{School of Physics and Microelectronics, Zhengzhou University,
             Zhengzhou, Henan 450001, China}\vspace{0.5cm}

\author{Qin-Tao Song}
\email{songqintao@zzu.edu.cn}
\affiliation{School of Physics and Microelectronics, Zhengzhou University,
             Zhengzhou, Henan 450001, China}\vspace{0.5cm}

\begin{abstract}
  Recently, the BESIII Collaboration reported a resonance $Y(2040)$  with  $M=2034\pm13\pm9$~MeV and $\Gamma=234\pm30\pm25$~MeV  in the process of $e^+e^-\rightarrow\omega\pi^0$.  In addition, new measurements with much higher precision for the $\rho(1900)$ and $\rho(2150)$ states are obtained by the BESIII and $BABAR$ Collaborations. In this work, we perform a systematic study on the mass spectrum of the excited  $\rho$ resonances  using the modified  Godfrey-Isgur model,
  and the strong decays of $Y(2040)$, $\rho(1900)$, and $\rho(2150)$  within the $^3P_0$ model.
  We find that  $Y(2040)$, $\rho(1900)$, and $\rho(2150)$ can be interpreted as the  $\rho(2^3D_1)$, $\rho(3^3S_1)$, and $\rho(4^3S_1)$ states, respectively. Meanwhile, the mass and strong decays of the $\rho(3^3D_1)$ state are predicted as well, which could be helpful to search for this state in future.

\end{abstract}

\date{Received 26 February 2021; accepted 15 July 2021}

\maketitle

\section{Introduction}
\label{sec1}

The study of low-energy properties of mesons such as mass spectra and decay modes is important to understand
the nonperturbative behaviors of QCD. Recently, there are abundant experimental data on the excited $\rho$ mesons, for example, the resonance $Y(2040)$ with  $M=2034\pm13\pm9$~MeV and $\Gamma=234\pm30\pm25$~MeV was observed in the process of $e^+e^-\rightarrow\omega\pi^0$ by the BESIII Collaboration~\cite{Ablikim:2020das}.
$Y(2040)$ could  be the same state as $\rho(2000)$ which is omitted from the summary table of the latest version of Particle Data Group (PDG)~\cite{Zyla:2020zbs},
because they have similar masses and decay widths~\cite{Anisovich:2001vt, Anisovich:2001jf, Anisovich:2000ut, Anisovich:2002su}. In addition, many new experimental measurements with much higher precision for $\rho(1900)$~\cite{Aubert:2006jq,Aubert:2007ym, Solodov:2011dn,CMD-3:2018kql}  and $\rho(2150)$~\cite{Lees:2012cj, Ablikim:2018iyx, BABAR:2019oes} are reported, especially recent measurements show that the decay width of $\rho(2150)$  is around $100$~MeV~\cite{Lees:2012cj, BABAR:2019oes, Ablikim:2018iyx} while a rather large width  $200-400$~MeV was obtained in previous measurements~\cite{Clegg:1989mp, Biagini:1990ze, Aubert:2007ef, Anisovich:2002su}. Thus, one can investigate the possible quark-model assignments for those states using these accurate experimental data.

In 1985, Godfrey and Isgur proposed the relativized quark model (GI model)~\cite{Godfrey:1985xj} which is widely used to investigate the mass spectrum of mesons, however, there appears a discrepancy between theoretical predictions and experimental values as more and more excited states were observed in the past decades.
As discussed in Ref.~\cite{Li:2009zu}, the linear confining potential $br$ will be screened and softened by the vacuum polarization effects at large distance,
thus the linear confining potential in a meson will be modified  by
the screening effects  at large distance,  which are induced by the quark-antiquark creation. In Refs.~\cite{Song:2015fha, Song:2015nia} the  screening effects are introduced into the GI model, as known the modified GI (MGI) model,  which gives a better description of  the mass spectrum for excited meson states.
In this work,  we shall adopt  the MGI model to estimate the mass spectrum of the excited $\rho$ mesons, based on which we give  assignments for $\rho(1900)$, $Y(2040)$, and $\rho(2150)$.
Furthermore, in order to check these assignments, we calculate the two-body OZI-allowed strong decays for $\rho(1900)$, $Y(2040)$, and $\rho(2150)$
within the $^3P_0$ model by employing the meson wave functions obtained from the MGI model.

The organization of this paper is as follows. In Sec.~\ref{sec2a}, a brief review of theoretical works and experiments is presented for $\rho(1900)$, $Y(2040)$, and $\rho(2150)$.
We introduce the MGI model and the $^3P_0$ model in Secs.~\ref{sec2}  and ~\ref{sec3}, respectively. The numerical results of mass spectrum and strong decays are shown in Sec.~\ref{sec4}. In Sec.~\ref{sec5} we give a summary of our results.

 \section{Current status of excited $\rho$ mesons}
\label{sec2a}

 \subsection{$\rho(1900)$}\label{rho1900}

The meson $\rho(1900)$ with $M=1870\pm 10$~MeV and $\Gamma=10\pm 5$~MeV was first observed by the FENICE Collaboration in the process of $e^+ e^-  \rightarrow N \bar{N}$~\cite{Antonelli:1996xn}, subsequently, a narrow dip structure, associated to the $\rho(1900)$ with $M=1910\pm 10$~MeV and $\Gamma=37\pm 13$~MeV, was also detected by the Fermilab  E687 Collaboration  through a study of the diffractive photoproduction of the $3 \pi^+ 3 \pi^-$ final state~\cite{Frabetti:2001ah,Frabetti:2003pw}. Recently, the existence of $\rho(1900)$ was further confirmed by the $BABAR$ Collaboration~\cite{Aubert:2006jq,Aubert:2007ym} and the CMD-3 Collaboration~\cite{Solodov:2011dn,CMD-3:2018kql}.
The measurements from $BABAR$ indicate that the widths of $\rho(1900)$ are $\Gamma=130\pm30$~MeV in  $e^+e^-\rightarrow3\pi^+3\pi^-\gamma$ and $\Gamma = 160\pm20$~MeV in $e^+e^-\rightarrow2(\pi^+\pi^-\pi^0)\gamma$~ \cite{Aubert:2006jq}.
The discrepancy of the decay widths  confuses our understanding of the $\rho(1900)$ nature.
A large decay width was obtained as $\Gamma=151^{+73}_{-75}$~MeV \cite{Matsinos:2020cft}   and  $\Gamma=186.8\pm 39.8$~MeV~\cite{Surovtsev:2008zza} by analyzing the related data.
Theoretically the assignment of $\rho(1900)$ as $\rho(3^3S_1)$  is supported by the study of Regge trajectories~\cite{Bugg:2012yt},  the $^3P_0$ model \cite{He:2013ttg}, and the effective Lagrangian approach \cite{ Wang:2020kte}.

\subsection{ $Y(2040)/\rho(2000)$} \label{rho2k}

A resonance with $J^{PC}=1^{--}$  was detected by analyzing the $p\bar{p}\rightarrow\pi^+\pi^-$ process at $M=1988$~MeV \cite{Hasan:1994he}, and it should be an excited  $\rho$ meson marked as $\rho(2000)$  since  the $P$ wave $\pi^+\pi^-$ can only have isospin $I=1$.
Later, the Crystal Barrel Collaboration found  a similar resonance  in the processes of $p\bar{p}\rightarrow \pi^+ \pi^-$, $p\bar{p}\rightarrow\omega\eta\pi^0$, and $p\bar{p}\rightarrow\omega\pi^0$~\cite{Anisovich:2001vt, Anisovich:2001jf, Anisovich:2000ut, Anisovich:2002su}. However, $\rho(2000)$ is omitted from the summary table of PDG~\cite{Zyla:2020zbs}, and needs to be further studied experimentally and theoretically.
Very recently, a resonance called  $Y(2040)$ was observed by the BESIII Collaboration with a significance of more than $10 \sigma$~\cite{Ablikim:2020das}, and the similar structure was also observed in the processes $J/\psi \rightarrow K^+ K^- \pi^0$~\cite{Ablikim:2019tqd} and $e^+e^-\to \eta' \pi^+\pi^-$~\cite{Aubert:2007ef,Ablikim:2020wyk}.
The resonance parameters of $Y(2040)$ are $M=2034\pm 13 \pm9 $~MeV and $\Gamma=234\pm30 \pm 25$~MeV~\cite{Ablikim:2020das}, respectively consistent with the $\rho(2000)$ mass and decay width  within the uncertainties~\cite{Anisovich:2001vt, Anisovich:2001jf, Anisovich:2000ut, Anisovich:2002su}.
Thus, $\rho(2000)$ and $Y(2040)$ are regarded as the same state in this work.

Based on the measurements with large uncertainties of the Crystal Barrel Collaboration~\cite{Anisovich:2000ut},
$\rho(2000)$  was assigned as $\rho(2^3D_1)$ in Ref.~\cite{He:2013ttg}  by studying its two-body OZI-allowed strong decay behaviors.
Thus, those accumulated abundant data with better accuracy on the resonance parameters provide an ideal lab to restudy the possible assignments of $\rho(2000)/Y(2040)$.

 \subsection{ $\rho(2150)$}

The $\rho(2150)$ was observed   more than 30 years ago~\cite{Bisello:1981sh,Atkinson:1985yx,Clegg:1989mp},
later it was  confirmed by the GAMS~\cite{Alde:1992wv,Alde:1994jm} and Crystal Barrel~\cite{Anisovich:1999xm,Anisovich:2002su, Anisovich:2000ut} Collaborations.
In 2007, the $BABAR$ Collaboration observed $\rho(2150)$  with $M=2150\pm 40\pm50  $~MeV and $\Gamma=350\pm 40\pm50 $~MeV in the process $e^+e^-\to f_1(1285) \pi^+\pi^-\gamma$~\cite{Aubert:2007ef}.
The early measurements of the $\rho(2150)$ width lie in the range of  200-400~MeV~\cite{Clegg:1989mp, Biagini:1990ze, Aubert:2007ef, Anisovich:2002su}, but  the recent measurements indicate  that the decay width of  $\rho(2150)$ is around 100~MeV as shown in Table~\ref{tab:2150}, where we list the measured resonance parameters  of $\rho(2150)$ since 2012.
There are four sets of parameters for $\rho(2150)$ in Ref.~\cite{BABAR:2019oes}, in which the different resonance parameters are obtained by separately and simultaneously fitting to the BESIII~\cite{Ablikim:2018iyx} and $BABAR$ \cite{Lees:2012cj, Lees:2013gzt, Lees:2018vvb}  datasets, respectively.
The parameters at the bottom of  Table~\ref{tab:2150} should be the most reliable among the four sets of parameters since this fit combines all decay modes in Refs.~\cite{Lees:2012cj, Lees:2013gzt, Lees:2018vvb, Ablikim:2018iyx}, and it is consistent with the latest results of the BESIII Collaboration\cite{Ablikim:2018iyx}.

\begin{table}
\begin{center}
\caption{ \label{tab:2150}   Mass and decay width of $\rho(2150)$ measured by experiments since 2012.   }
\begin{tabular}{cccc}
\hline\hline
 Mass (MeV)              & Width (MeV)          &Year   \\
\hline
 $2254\pm 22$ & $109\pm76$ \cite{Lees:2012cj} &2012\\
 $2239.2\pm 7.1 \pm 11.3$ &$139.8\pm12.3 \pm 20.6$ \cite{Ablikim:2018iyx}&2019 \\
 $2227 \pm 9 \pm 9 $   & $127 \pm 14 \pm 4$   \cite{BABAR:2019oes} &2020\\
 $2201\pm 19$              &   $70 \pm 38$     \cite{BABAR:2019oes}  &2020\\
 $2270 \pm 20 \pm 9$           &   $ 116^{+90}_{-60}\pm 50 $  \cite{BABAR:2019oes}  &2020    \\
 $2232 \pm 8 \pm 9$           &     $ 133\pm 14 \pm 4 $   \cite{BABAR:2019oes} &2020\\
\hline \hline
\end{tabular}
\end{center}
\end{table}

 The $\rho(2150)$ is a good candidate of $\rho(4^3S_1)$ according to the analysis of mass spectrum~\cite{Biagini:1990ze,Masjuan:2013xta,Bugg:2012yt}, and the results from the effective Lagrangian approach also support this assignment~\cite{Wang:2020kte}.
 In Ref.~\cite{He:2013ttg} the authors  studied the two-body strong decay behavior of  $\rho(2150)$ by using simple harmonic oscillator (SHO) functions as the wave functions of mesons, and a good description of $\rho(2150)$ can be obtained by considering it as  $\rho(4^3S_1)$, however, a large $\Gamma_{\rho(2150)}=230\pm50$~MeV~\cite{Anisovich:2002su} was adopted in the analysis. In this work we investigate the quark-model assignment for $\rho(2150)$ basing on more accurate measurements of the BESIII~\cite{Ablikim:2018iyx} and $BABAR$~\cite{BABAR:2019oes}  Collaborations.

\section{ Mass spectrum  }
\label{sec2}
In this section,  we will give a brief introduction of the GI model which was proposed by Godfrey and Isgur in 1985~\cite{Godfrey:1985xj}. The GI model plays an important role in studying the mass spectrum of mesons, especially for the low-lying states.
For high excited states, it is necessary to introduce the screening effects to the GI model, because the linear confining potential $br$ will be screened and softened by the vacuum polarization effects at large distance, as discussed in Refs.~\cite{Li:2009zu,Chao:1992et, Ding:1993uy},
The MGI model turns out to be able to give a better description of the mass spectra
for the higher radial and orbital excitations~\cite{Song:2015fha,Song:2015nia,Wang:2018rjg,Pang:2017dlw,Pang:2018gcn, Pang:2019ttv,Hao:2019fjg}.

\subsection{The GI model}
In GI model~\cite{Godfrey:1985xj}, the Hamiltonian of a meson includes the kinetic energy term and the effective potential term,
\begin{equation}\label{hamtn}
  \tilde{H}=\sqrt{m_1^2+\mathbf{p}^2}+\sqrt{m_2^2+\mathbf{p}^2}+\tilde{V}_{\mathrm{eff}}(\mathbf{p,r}),
\end{equation}
with
\begin{equation}
\tilde{V}_{\mathrm{eff}}(\mathbf{p,r})=\tilde{H}^\mathrm{conf}+\tilde{H}^\mathrm{hyp}+\tilde{H}^\mathrm{so},
\end{equation}
where $m_1$ and $m_2$ denote the masses of quark and antiquark, respectively,
$\tilde{V}_{\mathrm{eff}}(\mathbf{p,r})$ is the effective potential between quark and antiquark.
In the nonrelativistic limit, the effective potential can be simplified as
\begin{equation}
V_{\mathrm{eff}}(r)=H^{\mathrm{conf}}+H^{\mathrm{hyp}}+H^{\mathrm{so}}\label{1},
\end{equation}
with
\begin{align}
 H^{\mathrm{conf}}&=\Big[-\frac{3}{4}(c+br)+\frac{\alpha_s(r)}{r}\Big] \bm{F}_1\cdot\bm{F}_2  \nonumber\\
  &=S(r)+G(r)\label{3},\\ \nonumber \\
H^{\mathrm{hyp}}&=-\frac{\alpha_s(r)}{m_{1}m_{2}}\Bigg[\frac{1}{r^3}\Big(\frac{3\bm{S}_1\cdot\bm r \bm{S}_2\cdot\bm r}{r^2} -\bm{S}_1\cdot\bm{S}_2\Big)\nonumber\\
  &+\frac{8\pi}{3}\bm{S}_1\cdot\bm{S}_2\delta^3 (\bm r)\Bigg]  \bm{F}_1\cdot\bm{F}_2 ,\\ \nonumber\\
H^{\mathrm{so}}&=H^{\mathrm{so(cm)}}+H^{\mathrm{so(tp)}}, \label{3.2}
\end{align}
where $\langle \bm{F}_1\cdot\bm{F}_2 \rangle =-4/3$ for a meson.
The running coupling constant $\alpha_s(Q^2)$ depends on the energy scale $Q$ which is related to the relative momentum between quark and antiquark as $Q=|\vec{p}_1-\vec{p}_2|$. Since $\alpha_s(Q^2)$ is divergent at low $Q$ region, the authors of Ref.~\cite{Godfrey:1985xj} assume  that $\alpha_s(Q^2)$ saturates as $\alpha_s(Q^2=0)=\alpha_s^{\mathrm{critital}}$, and $\alpha_s^{\mathrm{critital}}$ is a parameter which is determined by fitting to the mass spectrum. $\alpha_s(r)$ is obtained from $\alpha_s(Q^2)$ by using Fourier transform, where $r$ is the relative distance between quark and antiquark.
$H^{\mathrm{conf}}$ reflects the spin-independent interaction, and it can be divided into two parts $S(r)$  and  $G(r)$. The linear confining potential $S(r)=b r+c$ plays an important role at large $r$, and the Coulomb-type potential $G(r)=-4 \alpha_s(r)/(3r)$ is dominant at small $r$.
$H^{\mathrm{hyp}}$ denotes the color-hyperfine interaction which can cause the mixing of different angular momenta, namely $ ^3L_J$ and $ ^3L_J^{\prime}$.
$H^{\mathrm{so}}$ is the spin-orbit interaction, which contains the color-magnetic term $H^{\mathrm{so(cm)}}$ and the Thomas-precession term $H^{\mathrm{so(tp)}}$.
The spin-orbit interaction will give rise to the  mixing between spin singlet $ ^1L_J$ and spin triplet $ ^3L_J$  if $m_1 \neq m_2$,
and the specific expression of $H^{\mathrm{so}}$  is  given in Ref.~\cite{Godfrey:1985xj}.

The GI model is a relativized quark model, and the relativistic effects are introduced by two main ways.
First, in the quark-antiquark scattering, the interactions should depend on both quark momentum
$\vec{p}_1$ and antiquark momentum $\vec{p}_2$, or a linear combination of them as $\vec{p}_1-\vec{p}_2$ and $\vec{p}_1+\vec{p}_2$, so they must be nonlocal interaction potentials as pointed out in Ref.~\cite{Godfrey:1985xj}. In order to take this effect into account,
a smearing {function} $\rho_{12} \left(\mathbf{r}-\mathbf{r}^{\prime}\right)$ is used to transform the basic potentials $G(r)$ and  $S(r)$ into the smeared ones $\tilde{G}(r)$ and  $\tilde{S}(r)$,
\begin{eqnarray}\label{sme}
\tilde{f}(r)=\int d^3r'\rho_{12}(\mathbf{r}-\mathbf{r'})f(r'),
\end{eqnarray}
and the smearing function is defined as
\begin{eqnarray}
& \rho_{12}\left(\mathbf{r}-\mathbf{r'}\right)=\frac{\sigma_{12}^3}{\pi ^{3/2}}e^{-\sigma_{12}^2\left(\mathbf{r}-\mathbf{r'}\right)^2}, \nonumber \\
&\sigma_{12}^2=\sigma_0^2\Bigg[\frac{1}{2}+\frac{1}{2}\left(\frac{4m_1m_2}{(m_1+m_2)^2}\right)^4\Bigg]+s^2\left(\frac{2m_1m_2}{m_1+m_2}\right)^2.
\label{si12}
\end{eqnarray}
 For a heavy-heavy $Q\bar{Q}$ meson system,  $\rho_{12}(\mathbf{r}-\mathbf{r'})$ will turn into delta function $\delta^3(\mathbf{r}-\mathbf{r'})$
as one increases the quark mass $m_Q$. In this case, one can obtain $\tilde{f}(r)=f(r)$,  which indicates  that  the relativistic effects can be neglected for a heavy-heavy $Q\bar{Q}$ meson.
However, the relativistic effects are important for heavy-light mesons  and light mesons, so it is necessary to adopt a relativized quark model such as the GI model to study the excited  $\rho$ mesons in this work.

Second,  the momentum-dependent factors are introduced to modify the effective potentials,
\begin{eqnarray}
 \tilde{G}(r)&\to \left(1+\frac{p^2}{E_1E_2}\right)^{1/2}\tilde{G}(r)\left(1 +\frac{p^2}{E_1E_2}\right)^{1/2},\\ \nonumber\\
 \frac{\tilde{V}_i(r)}{m_1m_2}&\to \left(\frac{m_1m_2}{E_1E_2}\right)^{1/2+\epsilon_i}\frac{\tilde{V}_i(r)}{m_1 m_2}\left(\frac{m_1 m_2}{E_1 E_2}\right)^{1/2+\epsilon_i},
\end{eqnarray}
where  $\tilde{G}(r)$ is the Coulomb-type potential, and $\tilde{V}_i(r)$ {represents} the contact, tensor, vector spin-orbit, and scalar spin-orbit terms as explained in Ref. \cite{Godfrey:1985xj}.  In the nonrelativistic limit, those momentum-dependent factors will become unity.

\subsection{The MGI model with screening effects}

 The following replacement  is often employed  to modify the linear confining potential $br$ in the quark model \cite{Chao:1992et, Ding:1993uy},
\begin{eqnarray}
V(r)=br\rightarrow V^{\mathrm{scr}}(r)=\frac{b(1-e^{-\mu r})}{\mu}.
\end{eqnarray}
If $r$ is small enough, we can have $V^{\mathrm{scr}}(r)=V(r)$. Therefore, this replacement will not affect the low-lying meson states.
The parameter $\mu$ is  related to the strength of the screening effects, and one can roughly understand that the screening effects begin to work from $r \sim 1/\mu$. The value of $\mu$ can be determined by fitting to the experimental measurements. Furthermore, the smeared potential \cite{Song:2015fha,Song:2015nia} can be obtained by using Eq. (\ref{sme}),
\begin{eqnarray}
\tilde V^{\mathrm{scr}}(r)&=&\frac{b}{\mu r}\Bigg[e^{\frac{\mu^2}{4 \sigma^2}+\mu r}\Bigg(\frac{1}{\sqrt{\pi}}\int_0^{\frac{\mu+2r\sigma^2}{2\sigma}}e^{-x^2}dx-\frac{1}{2}\Bigg)\nonumber\\
&&\times\frac{\mu+2r\sigma^2}{2\sigma^2}+r-e^{\frac{\mu^2}{4\sigma^2}-\mu r}\frac{\mu-2r\sigma^2}{2\sigma^2}\nonumber\\
&&\times\Bigg(\frac{1}{\sqrt{\pi}}\int_0^{\frac{\mu-2r\sigma^2}{2\sigma}}e^{-x^2}dx-\frac{1}{2}\Bigg)\Bigg], \label{Eq:pot}
\end{eqnarray}
 where $\sigma=\sigma_{12}$ defined in Eq.~(\ref{si12}).
 The MGI model is widely used in  Refs. \cite{Song:2015fha,Song:2015nia,Wang:2018rjg,Pang:2017dlw,Pang:2018gcn, Pang:2019ttv,Hao:2019fjg}, and it gives a better description of
 the mass spectrum  for both heavy-light mesons and light mesons.

 In this work, all the parameters involved in the MGI model
are listed in Table~\ref{tab:parameter}   followed Ref.~\cite{Pang:2018gcn}, where a systematic study is performed for the mass spectrum of the light mesons.
Here, we need to emphasize that there exist differences of parameters between relativistic quark models and nonrelativistic ones. For example, the mass parameters of quarks are roughly consistent as $m_u=m_d \sim 0.2 $ GeV and  $m_s \sim 0.4 $ GeV in relativistic quark models \cite{Godfrey:1985xj, Pang:2018gcn,Zeng:1994vj}, while
$m_u=m_d \sim 0.3 $ GeV and  $m_s \sim 0.6 $ GeV are often adopted by the nonrelativistic quark models \cite{Close:2005se,Zhong:2008kd}.
 The mass spectrum and wave functions of the mesons can be obtained by solving the Schr\"odinger  equation with  the Hamiltonian in Eq. (\ref{hamtn}).
 Furthermore,  the meson wave functions are used as inputs to investigate the subsequent strong decays for mesons.
 \begin{table}[htpb]
\begin{center}
\caption{ \label{tab:parameter}Parameter values in MGI model \cite{Pang:2018gcn}  }
\begin{tabular}{cccc}
\hline\hline
 Parameter                         & value  &  Parameter  &value\\
\hline
$m_u$~(GeV) &0.163 &$s$ &1.497 \\
$m_d$~(GeV) &0.163  &$\mu$~(GeV) &0.0635 \\
$m_s$~(GeV) &0.387&$\epsilon_{\rm c}$ &-0.138\\
 $b$~(GeV$^2$) &0.221&$\epsilon_{\rm{sov}}$ &0.157\\
  $c$~(GeV) &-0.240&$\epsilon_{\rm sos}$ &0.9726\\
$\sigma_0$~(GeV) &1.799&$\epsilon_{\rm t}$ &0.893\\
\hline\hline
\end{tabular}
\end{center}
\end{table}
\section{Two-body OZI-allowed strong decays }

\label{sec3}
 In addition to the mass spectrum, the decay widths are crucial  to identify the assignments for mesons.  Here, we give a brief introduction of the $^3P_0$ model which is widely used in studying two-body OZI-allowed strong decays of mesons~\cite{Li:2008we,Li:2008et,Xue:2018jvi,Wang:2017pxm,Pan:2016bac,Lu:2016bbk,Song:2014mha,Wang:2016krl,Pang:2014laa}.

\begin{figure}[htpb]
\includegraphics[scale=0.5]{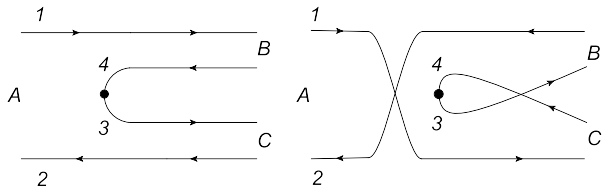}
\vspace{0.0cm}\caption{ Two possible diagrams contributing to the process $A\rightarrow BC$
in the $^3P_0$ model.}\label{fig:feiman}
\end{figure}

 The $^3P_0$ model was originally proposed by Micu \cite{Micu:1968mk} in 1968, and later it was  further developed by Le Yaouanc $et$ $al.$ \cite{LeYaouanc:1972vsx}. Hitherto, $^3P_0$ model has been considered as an effective tool to study two-body strong decays of hadrons, namely,  $A\rightarrow BC$.  In the decay process, a flavor-singlet and color-singlet quark-antiquark pair with $J^{PC}=0^{++}$ is created from the vacuum first, then, the created antiquark (quark) combines with the quark (antiquark)  in meson $A$ to form  meson $B$ ($C$) as shown in the left diagram of Fig.~\ref{fig:feiman}, besides, another similar decay mode is depicted  in the right diagram of Fig.~\ref{fig:feiman}.

The transition operator $T$ of the decay $A\rightarrow BC$ in the $^3P_0$ model is given by \cite{Blundell:1996as}
\begin{eqnarray}
T&=-3\gamma\sum\limits_m \langle 1m1-m|00\rangle\int d^3\boldsymbol{p}_3d^3\boldsymbol{p}_4\delta^3(\boldsymbol{p}_3+\boldsymbol{p}_4)\nonumber\\ \nonumber\\
&\times{\cal{Y}}^m_1\left(\frac{\boldsymbol{p}_3-\boldsymbol{p}_4}{2}\right
)\chi^{34}_{1,-m}\phi^{34}_0\omega^{34}_0b^\dagger_3(\boldsymbol{p}_3)d^\dagger_4(\boldsymbol{p}_4),
\end{eqnarray}
where $\boldsymbol{p}_3(\boldsymbol{p}_4$) is the momentum of the created quark (antiquark). $\gamma$ is a dimensionless parameter which stands for the strength of the quark-antiquark $q_3\bar{q}_4$ pair created from the vacuum, and it is often determined by fitting to the experimental data.
$\chi^{34}_{1,-m}$, $\phi^{34}_0$, and $\omega^{34}_0$ are spin, flavor, and color wave functions of the created quark-antiquark pair, respectively.

The helicity amplitude  ${\cal{M}}^{M_{J_A}M_{J_B}M_{J_C}}(\boldsymbol{P})$ of the decay process
 is defined with the help of the transition operator $T$,
\begin{eqnarray}
\begin{split}
\langle BC|T|A\rangle &=\delta^3(\boldsymbol{P}_A-\boldsymbol{P}_B-\boldsymbol{P}_C){\cal{M}}^{M_{J_A}M_{J_B}M_{J_C}}(\boldsymbol{P}),
\end{split}
\end{eqnarray}
where $|A\rangle$, $|B\rangle$, and $|C\rangle$ denote the mock meson states defined in Ref.~\cite{Hayne:1981zy}, and $\boldsymbol{P}$ is the momentum of meson $B$  in the center of mass frame.

Furthermore, one can express the partial wave amplitude ${\cal{M}}^{LS}(\boldsymbol{P})$ with helicity amplitude
for  the decay $A\rightarrow BC$ \cite{Jacob:1959at},
\begin{align}
{\cal{M}}^{LS}(\boldsymbol{P})&=\sum_{\renewcommand{\arraystretch}{.5}\begin{array}[t]{l}
\scriptstyle M_{J_B},M_{J_C},\\\scriptstyle M_S,M_L
\end{array}}\renewcommand{\arraystretch}{1}\!\!
\langle LM_LSM_S|J_AM_{J_A}\rangle \nonumber\\
&\times\langle J_BM_{J_B}J_CM_{J_C}|SM_S\rangle\nonumber\\
&\times\int d\Omega\,\mbox{}Y^\ast_{LM_L}{\cal{M}}^{M_{J_A}M_{J_B}M_{J_C}}
(\boldsymbol{P}).\label{pwave}
\end{align}
Finally, with the
relativistic phase space, the total width $\Gamma(A\rightarrow BC)$ can be
expressed in terms of  the partial wave amplitude squared \cite{Blundell:1996as},
\begin{eqnarray}
\Gamma(A\rightarrow BC)= \frac{\pi
|\boldsymbol{P}|}{4M^2_A}\sum_{LS}|{\cal{M}}^{LS}(\boldsymbol{P})|^2, \label{width1}
\end{eqnarray}
where $|\boldsymbol{P}|=\frac{\sqrt{[M^2_A-(M_B+M_C)^2][M^2_A-(M_B-M_C)^2]}}{2M_A}$, $M_A$, $M_B$, and $M_C$ are the masses of the mesons $A$, $B$, and $C$, respectively.

\section{Numerical RESULTS}
\label{sec4}

\subsection{Mass spectrum analysis}

\begin{figure}[htpb]
\includegraphics[scale=0.4]{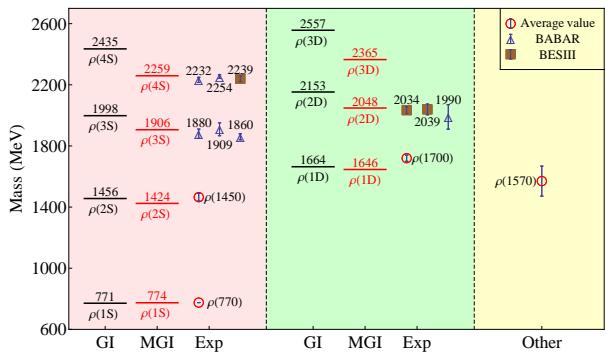}
\vspace{0.0cm}
\caption{The theoretical and experimental  masses of $\rho$ mesons, the black solid and red solid lines denote   the predictions from the GI model and MGI model, respectively.}\label{fig:zhiliang}
\end{figure}

Here, we adopt both the MGI and GI models to calculate the mass spectrum for excited $\rho$ mesons. The parameters involved in the
GI model are taken from  Ref.~\cite{Godfrey:1985xj}.
In Table~\ref{tab:mass}, the numerical results are shown together with recent experimental measurements.
Here,  we need to mention that, in 2008, the $BABAR$ Collaboration announced an enhancement around 1.57 GeV \cite{Aubert:2007ym} which is named as $\rho(1570)$ shown in Fig.~\ref{fig:zhiliang},  however, this state has not been confirmed by other experiments since 2008. Therefore,  we will not discuss it in this work.
The mass differences between the MGI model and the GI model become larger for higher excited $\rho$ mesons as shown in Fig.~\ref{fig:zhiliang}. For example, the mass differences are as large as $100$~MeV for $\rho(3^3S_1)$,
$\rho(2^3D_1)$, and
 $\rho(4^3S_1)$.
 Comparing  the MGI model predictions with the measured masses, one can regard $\rho(1900)$, $Y(2040)$, and $\rho(2150)$ as possible candidates of $\rho(3^3S_1)$,
$\rho(2^3D_1)$, and
 $\rho(4^3S_1)$, respectively.

\begin{table}[htpb]
\begin{center}
\caption{ \label{tab:mass}The predicted  and measured masses of $\rho$ mesons.}
\begin{tabular}{cccc}
\hline\hline
 State                         & MGI (MeV)            & GI (MeV) &Exp (MeV) \\
\hline
$\rho(1^3S_1)$ &774  &771 &$775.26\pm0.25$~\cite{Zyla:2020zbs} \\
$\rho(2^3S_1)$ &1424  &1456&$1465\pm25$~\cite{Zyla:2020zbs} \\
$\rho(3^3S_1)$ &1906  &1998    &$1909\pm17\pm25$ \cite{Aubert:2007ym} \\
              &       &           &$1880\pm30$\cite{Aubert:2006jq} \\
              &       &            &$1860\pm20$ \cite{Aubert:2006jq} \\
$\rho(4^3S_1)$ &2259&2435   &$2239.2\pm 7.1 \pm 11.3$ \cite{Ablikim:2018iyx} \\
               &     &          &    $2232 \pm 8 \pm 9$     \cite{BABAR:2019oes}  \\
               &     &           &  $2254\pm 22$  \cite{Lees:2012cj}  \\
        $\rho(5^3S_1)$ &2542 &2817     &$\cdots$\\
$\rho(1^3D_1)$&1646&1664     &$1720\pm20$~\cite{Zyla:2020zbs}\\
$\rho(2^3D_1)$&2048&2153    &$2034\pm13\pm9$ \cite{Ablikim:2020das}\\
             &     &           & $2039 \pm 8 ^{+36}_{-18}$\cite{Ablikim:2019tqd}  \\
             &      &          &$1990 \pm 80$   \cite{Aubert:2007ef}\\
$\rho(3^3D_1)$ &2365 &2557     &$\cdots$\\
$\rho(4^3D_1)$ &2624 &2915     &$\cdots$\\
\hline\hline
\end{tabular}
\end{center}
\end{table}

\subsection{Decay behavior analysis}

 It should be emphasized that the mass alone is insufficient to identify those assignments, and the decay behaviors need to be analyzed. In this work, we employ the $^3P_0$ model with  the realistic meson wave functions obtained from the MGI model to evaluate the decay widths of $\rho(1900)$, $Y(2040)$, and $\rho(2150)$. In this case, only the parameter $\gamma$ is unknown in the $^3P_0$ model. Since  $\rho(1900)$, $Y(2040)$, and $\rho(2150)$ are light mesons, we can assume they share the same $\gamma$  with other excited light mesons with $J^{PC}=1^{--}$. We obtain $\gamma= 6.57$ for $u\bar{u}/d\bar{d}$ pair creation by fitting to the total widths of  $\rho(1700)$, $K^*(1680)$, $\omega(1650)$, $\rho(1450)$, $K^*(1410)$, $\phi(1680)$, and $\omega(1420)$ mesons.  As for the $s\bar{s}$ pair creation in the decay process, the $\gamma$ value is multiplied  by a factor $m_u/m_s$.

\begin{table}[htpb]
\begin{center}
\caption{ \label{tab:2D}Decay widths of $Y(2040)$ as the $\rho(2^3D_1)$ (in MeV), the initial mass is set to be 2034~MeV and the masses of all the final states are taken from PDG~\cite{Zyla:2020zbs}.}
\begin{tabular}{c|cc|cc}
\hline\hline
 Channel                 & Mode      &$\rho(2^3D_1)$            & Mode           &$\rho(2^3D_1)$   \\
\hline
 $1^-\rightarrow 0^-0^-$      & $\pi\pi$     & 19.77  & $KK$  &0.32 \\
                              & $\pi\pi(1300)$  & 14.81 & $KK(1460)$  &0.30    \\
                              & $\pi\pi(1800)$  & 1.28 &    &     \\
                                                         \hline
  $1^-\rightarrow 0^-1^-$     &$\pi\omega$  &6.31  &$\rho\eta'$ &0.013         \\
                              &$\rho\eta$  &2.17    &$KK^*$  &0.015     \\
                              &$\omega(1420)\pi$ &6.68 &$\omega(1650)\pi$  &0.14  \\
                              &$KK^*(1410)$  &0.57   &$\rho(1450)\eta$ & 0.51   \\
                               \hline
  $1^-\rightarrow 1^-1^-$     &$\rho\rho$   &36.38   &$K^*K^*$  &0.13\\
                               \hline
  $1^-\rightarrow 0^-1^+$    &$a_{1}(1260)\pi$   &26.60   &$h_{1}(1170)\pi$ &35.20\\
                              &$KK_{1}(1400)$  &0.098    &$b_{1}(1235)\eta$  &6.23 \\
                               &$KK_{1}(1270)$  &0.19&  &  \\
                                 \hline
  $1^-\rightarrow 0^-2^+$     &$a_{2}(1320)\pi$  &9.76   &$KK_{2}^*(1430)$ &0.027\\
                               \hline
  $1^-\rightarrow 0^- 2^-$            &$\pi\pi_{2}(1670)$ &39.15&   & \\
   \hline
  $1^-\rightarrow 0^- 3^-$            &$\pi\omega_{3}(1670)$ &0.19 &  &  \\
   \hline
  $1^-\rightarrow 1^- 1^+$            &$b_{1}(1235)\rho$  &15.86 &$a_{1}(1260)\omega$  &5.20 \\
                                       \hline
Total width    &\multicolumn{4}{c}{227.91}\\
\hline
Experiment    &\multicolumn{4}{c}{$234\pm30\pm25$~\cite{Ablikim:2020das}}\\
\hline\hline
\end{tabular}
\end{center}
\end{table}

\begin{figure}[htpb]
\includegraphics[scale=0.75]{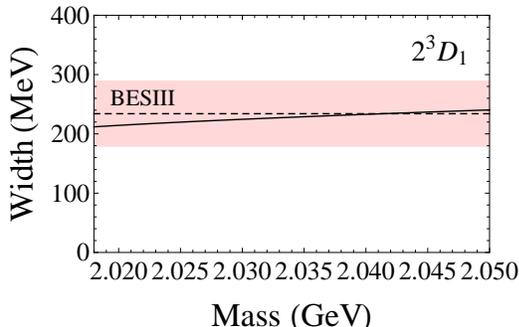}
\vspace{-0.5cm}
\caption{The dependence of the decay width on the mass of $Y(2040)$ as $\rho(2^3D_1)$, and the pink error
band indicates the measured width for the $Y(2040)$ from the BESIII Collaboration~\cite{Ablikim:2020das}}\label{fig:decay2d}
\end{figure}

The partial widths and total width of  $Y(2040)$ as  $\rho(2^3D_1)$ are listed in Table~\ref{tab:2D}.
The total width is expected to be 227.91~MeV, in good agreement with the recent BESIII measurement of
$\Gamma_{Y(2040)}=234\pm30\pm25$~MeV~\cite{Ablikim:2020das}.
Besides, the predicted branching ratio is  $\Gamma_{K^+ K^-}/ \Gamma \approx 0.14\% $ under the assignment of $\rho(2^3D_1)$, also consistent  with the BESIII measurement $\Gamma_{K^+ K^-}/ \Gamma=0.1\%- 0.2 \%$~\cite{Ablikim:2019tqd}.
The main decay modes are $\pi \pi$, $\pi \pi(1300)$, $\rho\rho$, $\pi a_1(1260)$, $a_2(1320)\pi$, $\pi\pi_2(1670)$, $b_1(1235) \rho$,  and  $\pi h_1(1170)$. The background of $\pi\pi$ is complicated,
and it is difficult to identify new resonances. Apart from  $\pi \pi$ and $a_2(1320)\pi$,
there exists at least one broad state in other main channels,
so it is hard to reconstruct the two-body processes involving a broad state experimentally.
Therefore, $a_2(1320)\pi$ should be a good channel for investigating $Y(2040)$, and the decay width is 9.76 MeV which is comparable with the decay width 6.31 MeV of the observed mode $\pi \omega$.
The dependence of the total width on the mass of the initial state is shown in Fig.~\ref{fig:decay2d}, and the pink error band indicates the decay width $\Gamma_{Y(2040)}=234\pm30\pm25$~MeV~\cite{Ablikim:2020das}.
In the mass range of the experimental value $M_{Y(2040)}=2034\pm13\pm9$~MeV, the predicted total width is always in agreement with the measurement. Thus, we can have a nice description of $Y(2040)$ on the mass spectrum and decay behaviors under the assignment of $\rho(2^3D_1)$.

\begin{table}[htpb]
\begin{center}
\caption{ \label{tab:1900}Decay widths of $\rho(1900)$ and $\rho(2150)$ as $\rho(3^3S_1)$ and $\rho(4^3S_1)$ states (in MeV), the initial state masses are set to be 1880~MeV and 2232~MeV, respectively. The masses of all the final states are taken from PDG~\cite{Zyla:2020zbs}}
\begin{tabular}{c|ccc}
\hline\hline
 Channel                 & Mode      &$\rho(1900)$    &$\rho(2150)$    \\
\hline
 $1^-\rightarrow 0^-0^-$      & $\pi\pi$        &10.89   &6.18  \\
                              & $\pi\pi(1300)$    &14.52    &15.96  \\
                               & $\pi\pi(1800)$    &$\cdots$     &10.07 \\
  $ $                         & $KK$  &0.34        &0.032\\
                               & $KK(1460)$   &$\cdots$       & 0.12  \\
                                \hline
  $1^-\rightarrow 0^-1^-$     &$\pi\omega$    &21.07    &5.41  \\
                              &$\rho\eta$    &4.63     &0.82 \\
                              &$\omega(1420)\pi$    &29.61      &18.92 \\
                              &$\omega(1650)\pi$   &0.017        & 0.15 \\
                              &$KK^*$    &0.27     & 0.0098\\
                              &$KK^*(1410)$   &$\cdots$        & 0.40\\
                              &$\rho\eta'$   &0.13      &0.11 \\
                              &$\rho(1450)\eta$   &$\cdots$    &0.59 \\
                               &$\rho\eta(1295)$   &$\cdots$  &0.19 \\
                              &$\pi(1300)\omega$   &$\cdots$     & 0.14\\
 \hline
  $1^-\rightarrow 1^-1^-$     &$\rho\rho$   &0.011      &2.19 \\
                              &$K^*K^*$   &0.11       & 0.41\\
                               \hline
  $1^-\rightarrow 0^-1^+$    &$a_{1}(1260)\pi$     &10.16        & 7.73\\
                              &$h_{1}(1170)\pi$    &8.61      & 8.40 \\
                              &$KK_{1}(1400)$    &$\cdots$      &0.35  \\
                               &$KK_{1}(1270)$   &$0.94$    & 0.32  \\
                                &$b_{1}(1235)\eta$   &1.76   & 1.26\\
                                 \hline
  $1^-\rightarrow 0^-2^+$     &$a_{2}(1320)\pi$    &22.36   & 14.71 \\
   \hline
  $1^-\rightarrow 0^- 2^-$            &$\pi\pi_{2}(1670)$  &0.067   & 3.44 \\
   \hline
  $1^-\rightarrow 0^- 3^-$            &$\pi\omega_{3}(1670)$   &0.0043      &2.24  \\
   \hline
  $1^-\rightarrow 1^- 1^+$            &$b_{1}(1235)\rho$   &$\cdots$      &5.88\\
                                      &$f_{1}(1285)\rho$ &$\cdots$   &4.49 \\
                                      &$a_{1}(1260)\omega$   &$\cdots$    & 5.86\\
                                       \hline
  $1^-\rightarrow 1^- 2^+$            &$f_{2}(1270)\rho$   &$\cdots$     &2.30 \\
                                      &$a_{2}(1320)\omega$   &$\cdots$   &3.03 \\
                                       \hline
   $1^-\rightarrow 0^- 4^+$            &$a_{4}(1970)\pi$   &$\cdots$    &0.0048\\
    \hline
 \multicolumn{2}{c|}{Total width}    &125.51 &121.72    \\
 \hline
 \multicolumn{2}{c|}{Experiment}   &$130\pm30$ \cite{Aubert:2006jq}  &$ 133\pm 14 \pm 4 $~\cite{BABAR:2019oes}\\
\hline\hline
\end{tabular}
\end{center}
\end{table}

\begin{figure}[htpb]
\includegraphics[scale=0.75]{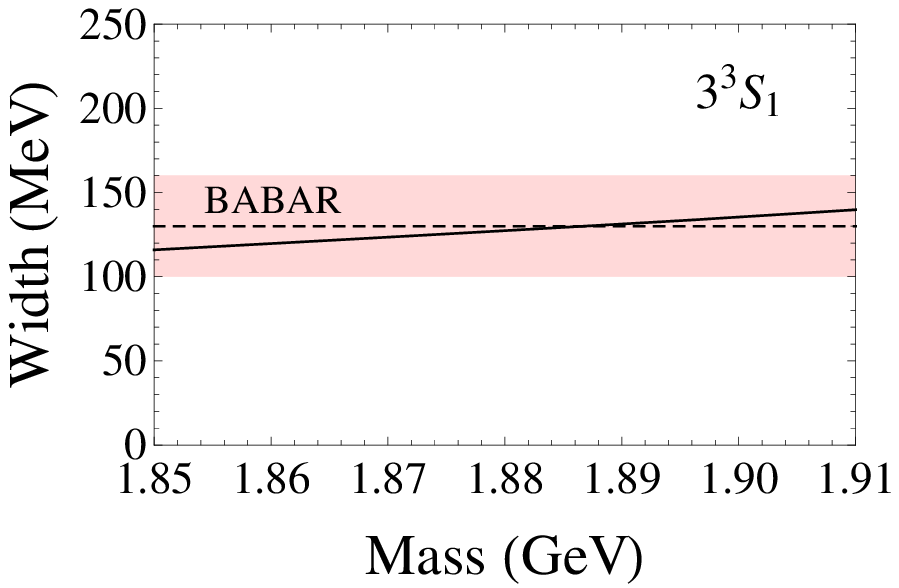}
\vspace{0.0cm}
\caption{The dependence of the decay width on the mass of $\rho(1900)$ as $\rho(3S)$, and the pink error band indicates the measured width for the $\rho(1900)$ from the $BABAR$ Collaboration~\cite{Aubert:2006jq}.}\label{fig:decay19}
\end{figure}

 The decay widths of $\rho(1900)$ as $\rho(3^3S_1)$ are shown in Table~\ref{tab:1900}, and the main decay modes are $\pi\pi$, $\pi a_1(1260)$, $\pi a_2(1320)$, $\pi\pi(1300)$, $\omega(1420)\pi$, and $\pi\omega$.
Considering the experimental conditions, $\pi a_2(1320)$ and $\pi\omega$ should be the best channels for investigating $\rho(1900)$.
  The predicted total width of $\rho(1900)$ as $\rho(3^3S_1)$ is 125.51~MeV which agrees with the $BABAR$ measurement of $\Gamma_{\rho(1900)}=130\pm30$~MeV~\cite{Aubert:2006jq} very well.
As we discussed above, there exists a discrepancy for  the  $\rho(1900)$ width among different experiments.
Therefore, further precise measurements for the $\rho(1900)$ width are necessary to pin down the $\rho(3^3S_1)$ assignment.
In Fig.~\ref{fig:decay19}, we also show the dependence of the total width on the mass of $\rho(1900)$ as $\rho(3^3S_1)$, the pink error band indicates  the uncertainties of the measured width.
In the mass range of the experimental value $M_{\rho(1900)}=1880\pm30$~MeV~\cite{Aubert:2006jq}, the predicted total width is still in agreement with the $BABAR$ value $\Gamma_{\rho(1900)}=130\pm30$~MeV.

\begin{figure}[htpb]
\includegraphics[scale=0.75]{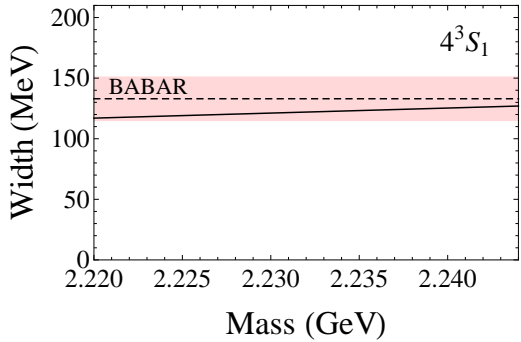}
\vspace{0.0cm}\caption{The dependence of the decay width on the mass of $\rho(2150)$ as  $\rho(4^3S_1)$, and the pink error band indicates the measured width for the $\rho(2150)$ from the $BABAR$ Collaboration~\cite{BABAR:2019oes}.}\label{fig:decay}
\end{figure}

The decay widths of $\rho(2150)$ as $\rho(4^3S_1)$ are shown in Table~\ref{tab:1900}.
The total width is predicted to be 121.72~MeV, which is in good agreement with
the BESIII measurement of $139.8\pm12.3 \pm 20.6$~MeV~\cite{Ablikim:2018iyx} and the $BABAR$ measurement of $ 133\pm 14 \pm 4 $~MeV~\cite{BABAR:2019oes}.
 We can see that the main decay channels are $\pi\pi(1300)$, $\pi\pi(1800)$, $\pi\omega(1420)$, and $\pi a_2(1320)$,
among them $\pi a_2(1320)$ should be the easiest one to be measured by experiment as we discussed for $Y(2040)$.
In Fig.~\ref{fig:decay}, we show the dependence of the total width on the mass of $\rho(2150)$.
In the mass range of the experimental value $M_{\rho(2150)}=2232 \pm 8 \pm 9$ ~MeV~\cite{BABAR:2019oes}, the predicted total width is still in agreement with the $BABAR$ value $ 133\pm 14 \pm 4 $ MeV.
Therefore, $\rho(2150)$ is a very good candidate of $\rho(4^3S_1)$.

 We also predict the decay behavior for $\rho(3^3D_1)$ state with the initial mass $M=2365$~MeV obtained in the MGI model, as listed  in Table~\ref{tab:3D}.
The $\rho(3^3D_1)$ width is predicted to be $204.89$~MeV, and
the main decay modes are $\rho\rho$, $\pi h_1(1170)$, $\pi \pi_2(1670)$, and $\pi a_1(1640)$. It should be stressed that the state
$\rho(2270)$ with a mass of $2265\pm40$~MeV and a width of  $325\pm 80$~MeV was reported by the Crystal Barrel Collaboration~\cite{Anisovich:2000ut,Anisovich:2002su} , but is not further confirmed by other experiments.  In Ref.~\cite{He:2013ttg}, the $\rho(2270)$ was regarded as the candidate of $\rho(3^3D_1)$.  Our predictions on the mass and width for the $\rho(3^3D_1)$ do not favor the assignment of $\rho(2270)$ as $\rho(3^3D_1)$.

\begin{table}[htpb]
\begin{center}
\caption{ \label{tab:3D}The decay widths of $\rho(3^3D_1)$ (in MeV),  the initial state mass is set to be 2365 MeV and the masses of all the final states are taken from PDG~\cite{Zyla:2020zbs}.}
\begin{tabular}{c|cc|cc}
\hline\hline
 Channel                 & Mode      &$\rho(3^3D_1)$            & Mode           &$\rho(3^3D_1)$   \\
\hline
 $1^-\rightarrow 0^-0^-$      & $\pi\pi$     & 12.59  & $KK$  &0.029 \\
                              & $\pi\pi(1300)$  & 14.34 & $KK(1460)$  &0.097    \\
                               & $\pi\pi(1800)$  &10.45   &  &    \\
  \hline
  $1^-\rightarrow 0^-1^-$     &$\pi\omega$  &2.28         &$KK^*(1680)$ & 0.043 \\
                              &$\rho\eta$  &0.56       &$\rho(1700)\eta$ & 0.082 \\
                              &$\omega(1420)\pi$ &3.73   &$\rho\eta(1475)$ & 0.090  \\
                              &$\omega(1650)\pi$  &0.59   &$\omega\pi(1300)$ & 1.17    \\
                              &$KK^*$  &0.013  &$\rho\eta(1295)$ & 0.70  \\
                              &$KK^*(1410)$  &0.031   &$\rho(1450)\eta$ & 0.89   \\
                              &$\rho\eta'$ &0.0020  & &     \\
                               \hline
  $1^-\rightarrow 1^-1^-$     &$\rho\rho$   &24.88   &$K^*K^*$  &0.039\\
                                &$\rho\rho(1450)$   &13.15    &$K^*K^*(1410)$  &0.073   \\
                               \hline
  $1^-\rightarrow 0^-1^+$    &$a_{1}(1260)\pi$   &13.08   &$b_{1}(1235)\eta'$  &0.25    \\
                              &$h_{1}(1170)\pi$   &17.54   &$\pi a_1(1640)$  &16.38 \\
                              &$KK_{1}(1400)$  &0.11   &$b_{1}(1235)\eta$  &2.94 \\
                               &$KK_{1}(1270)$  &0.051&  &  \\
                                 \hline
  $1^-\rightarrow 0^-2^+$     &$a_{2}(1320)\pi$  &7.92 &$a_{2}(1700)\pi$  &4.98 \\
                              &$KK_{2}^*(1430)$ &0.0016&  & \\
                               \hline
  $1^-\rightarrow 0^- 2^-$            &$\pi\pi_{2}(1670)$ &17.69 &$\pi\eta_{2}(1645)$ &11.51\\
                                      &$KK_2(1770)$ &0.019 &$KK_2(1820)$ &0.0018\\
   \hline
  $1^-\rightarrow 0^- 3^-$            &$\pi\omega_{3}(1670)$ &1.12 &$\eta\rho_{3}(1690)$ &0.11 \\
                                      &$KK_{3}(1780)$ &0.0007 &   &  \\
   \hline
  $1^-\rightarrow 1^- 1^+$            &$b_{1}(1235)\rho$  &3.35 &$K_1(1400)K^*$&0.077\\
                                      &$a_{1}(1260)\omega$  &2.17  &$K_1(1270)K^*$&0.25\\
                                      &$\rho f_1(1285)$&1.18 &$\rho f_1(1420)$&0.17\\
                                      &$\rho h_1(1170)$&4.52&  & \\
                                       \hline
     $1^-\rightarrow 2^+ 1^-$          &$\rho f_2(1270)$&7.71&$K^*K_2^*(1430)$&0.022\\
                                         &$\omega a_2(1320)$&5.33&  & \\
                                         \hline
         $1^-\rightarrow 0^- 4^+$        & $a_4(1970)\pi$&0.11&  &   \\
         \hline
           $1^-\rightarrow 1^- 0^+$          &$a_0(1450)\omega$&0.46&  &   \\
           \hline

Total width    &\multicolumn{4}{c}{204.89}\\
\hline\hline
\end{tabular}
\end{center}
\end{table}

\section{SUMMARY AND CONCLUSION}
\label{sec5}
In this work, we perform a systematic study on the mass spectrum and decay properties for $Y(2040)$, $\rho(1900)$, and $\rho(2150)$.  The mass spectrum of the excited $\rho$ mesons is predicted within the MGI model where the screening effects are taken into account. Moreover,  the decay behaviors are calculated in the $^3P_0$ model with the meson wave functions obtained from the MGI model.
We draw the following conclusions by comparing the recent precise experimental measurements with our theoretical results.

 \begin{enumerate}
  \item The screening effects play an important role in studying  the masses of $Y(2040)$, $\rho(1900)$, $\rho(2150)$, and $\rho(3^3D_1)$,  and mass gaps around 100~MeV appear when we compare the MGI model predictions with the ones of the GI model.

\item  The newly observed state $Y(2040)$ should be the same state as $\rho(2000)$ which is omitted in the summary table of PDG~\cite{Zyla:2020zbs}, since they share the similar resonance parameters~\cite{Aubert:2007ef, Ablikim:2019tqd, Ablikim:2020das} .

\item
The $Y(2040)$, $\rho(1900)$, and $\rho(2150)$ can be interpreted as the  $\rho(2^3D_1)$, $\rho(3^3S_1)$, and $\rho(4^3S_1)$ states, respectively.

 \end{enumerate}

This work is helpful for us  to  reveal the inner structure of  $Y(2040)$, $\rho(1900)$, and $\rho(2150)$,  which is crucial to understand  the $\rho$ meson family. In addition, we expect that more and more experimental measurements will be released  in  future, especially the decay width of $\rho(1900)$ needs to be measured precisely to pin down the $\rho(3^3S_1)$  assignment.

Apart from the phenomenological models employed in this work, much progress has been made on mass spectrum and strong decays of  the $\rho$ mesons by using lattice QCD in the past decade. For example, masses  are estimated in Ref.~\cite{Dudek:2010wm} for a few excited $\rho$ mesons, and the authors of Ref.~\cite{Johnson:2020ilc} also compute the decay widths  $\pi \rho$ and $\pi \omega$ for $\rho(1450)$  and $\rho(1700)$.
We expect that more and more results will be released by  lattice QCD in the near future, and the  predictions of this work can be checked for the excited $\rho$ mesons.

\section*{Acknowledgements}

This work is partly supported by the National Natural Science Foundation of China under Grants No. 12005191,
the Key Research Projects of Henan Higher Education Institutions under No. 20A140027, Training Plan for Young Key Teachers in Higher Schools in Henan Province (2020GGJS017), the Academic Improvement Project of Zhengzhou University, the Fundamental Research Cultivation Fund for Young Teachers of Zhengzhou University (JC202041042).


\begin{thebibliography}{99}

\bibitem{Ablikim:2020das}
M.~Ablikim \textit{et al.} [BESIII],
``Observation of a resonant structure in $e^{+}e^{-} \to \omega\eta$ and another in $e^{+}e^{-} \to \omega\pi^{0}$ at center-of-mass energies between 2.00 and 3.08 GeV,''
Phys. Lett. B \textbf{813} (2021), 136059.



\bibitem{Zyla:2020zbs}
P.~A.~Zyla \textit{et al.} [Particle Data Group],
``Review of Particle Physics,''
PTEP \textbf{2020} (2020) no.8, 083C01.


\bibitem{Anisovich:2002su}
A.~V.~Anisovich, C.~A.~Baker, C.~J.~Batty, D.~V.~Bugg, L.~Montanet, V.~A.~Nikonov, A.~V.~Sarantsev, V.~V.~Sarantsev and B.~S.~Zou,
``Combined analysis of meson channels with $I = 1$, $C = -1$ from 1940 to 2410 MeV,''
Phys.Lett.B \textbf{542} (2002), 8-18.


\bibitem{Anisovich:2001vt}
A.~V.~Anisovich, C.~A.~Baker, C.~J.~Batty, D.~V.~Bugg, V.~A.~Nikonov, A.~V.~Sarantsev, V.~V.~Sarantsev and B.~S.~Zou,
``Resonances in $\bar{p} p  \to \omega \eta \pi^0$ in the mass range 1960~MeV to 2410~MeV,''
Phys. Lett. B \textbf{513} (2001), 281-291.


\bibitem{Anisovich:2001jf}
A.~V.~Anisovich, C.~A.~Baker, C.~J.~Batty, D.~V.~Bugg, V.~A.~Nikonov, A.~V.~Sarantsev, V.~V.~Sarantsev and B.~S.~Zou,
``Resonances in $\bar{p} p  \to \omega  \pi^0$ in the mass range 1960~MeV to 2410~MeV,''
Phys. Lett. B \textbf{508} (2001), 6-16.

\bibitem{Anisovich:2000ut}
A.~V.~Anisovich, C.~A.~Baker, C.~J.~Batty, D.~V.~Bugg, C.~Hodd, H.~C.~Lu, V.~A.~Nikonov, A.~V.~Sarantsev, V.~V.~Sarantsev and B.~S.~Zou,
``$I = 0~C = +1$ mesons from 1920 to 2410 MeV,''
Phys. Lett. B \textbf{491} (2000), 47-58.


\bibitem{Aubert:2006jq}
B.~Aubert \textit{et al.} [BaBar],
``The $e^+e^-$ $ \to 3(\pi^+ \pi^-)$, $2(\pi^+ \pi^- \pi^0)$ and $K^+ K^- 2(\pi^+ \pi^-)$ cross sections at center-of-mass energies from production threshold to 4.5-GeV measured with initial-state radiation,''
Phys. Rev. D \textbf{73} (2006), 052003.

\bibitem{Aubert:2007ym}
B.~Aubert \textit{et al.} [BaBar],
``Measurements of $e^{+} e^{-} \to K^{+} K^{-} \eta$, $K^{+} K^{-} \pi^0$ and $K^0_{s} K^\pm \pi^\mp$ cross sections using initial state radiation events,''
Phys. Rev. D \textbf{77} (2008), 092002.

\bibitem{Solodov:2011dn}
E.~P.~Solodov [CMD-3],
``First results from the CMD3 Detector at the VEPP2000 Collider,''
[arXiv:1108.6174 [hep-ex]].

\bibitem{CMD-3:2018kql}
R.~R.~Akhmetshin \textit{et al.} [CMD-3],
``Observation of a fine structure in $e^+ e^- \to$ hadrons production at the nucleon-antinucleon threshold,''
Phys. Lett. B \textbf{794} (2019), 64-68.


\bibitem{Ablikim:2018iyx}
M.~Ablikim \textit{et al.} [BESIII],
``Measurement of $e^{+} e^{-} \rightarrow K^{+} K^{-}$ cross section at $\sqrt{s} = 2.00 - 3.08$ GeV,''
Phys. Rev. D \textbf{99} (2019) no.3, 032001.


\bibitem{BABAR:2019oes}
J.~P.~Lees \textit{et al.} [BaBar],
``Resonances in $e^+e^-$ annihilation near 2.2 GeV,''
Phys. Rev. D \textbf{101} (2020) no.1, 012011.


\bibitem{Lees:2012cj}
J.~P.~Lees \textit{et al.} [BaBar],
``Precise Measurement of the $e^+ e^- \to \pi^+\pi^- (\gamma)$ Cross Section with the Initial-State Radiation Method at BABAR,''
Phys. Rev. D \textbf{86} (2012), 032013.

\bibitem{Biagini:1990ze}
M.~E.~Biagini, S.~Dubnicka, E.~Etim and P.~Kolar,
``Phenomenological evidence for a third radial excitation of $\rho(770)$,''
Nuovo Cim. A \textbf{104} (1991), 363-370.


\bibitem{Clegg:1989mp}
A.~B.~Clegg and A.~Donnachie,
``$\rho^\prime$s in 6 $\pi$ States From Materialization of Photons,''
Z. Phys. C \textbf{45} (1990), 677.




\bibitem{Aubert:2007ef}
B.~Aubert \textit{et al.} [BaBar],
``The $e^+ e^-\to 2(\pi^+ \pi^-)$  $\pi^0, 2(\pi^+ \pi^-) \eta$, $K^+ K^- \pi^+ \pi^- \pi^0$ and $K^+ K^- \pi^+ \pi^- \eta$ Cross Sections Measured with Initial-State Radiation,''
Phys. Rev. D \textbf{76} (2007), 092005
[erratum: Phys. Rev. D \textbf{77} (2008), 119902].



\bibitem{Godfrey:1985xj}
S.~Godfrey and N.~Isgur,
``Mesons in a Relativized Quark Model with Chromodynamics,''
Phys. Rev. D \textbf{32} (1985), 189-231.

\bibitem{Li:2009zu}
B.~Q.~Li and K.~T.~Chao,
``Higher Charmonia and $X$,~$Y$,~$Z$ states with Screened Potential,''
Phys. Rev. D \textbf{79} (2009), 094004.

\bibitem{Song:2015nia}
Q.~T.~Song, D.~Y.~Chen, X.~Liu and T.~Matsuki,
``Charmed-strange mesons revisited: mass spectra and strong decays,''
Phys. Rev. D \textbf{91} (2015), 054031.

\bibitem{Song:2015fha}
Q.~T.~Song, D.~Y.~Chen, X.~Liu and T.~Matsuki,
``Higher radial and orbital excitations in the charmed meson family,''
Phys. Rev. D \textbf{92} (2015) no.7, 074011.

\bibitem{Antonelli:1996xn}
A.~Antonelli \textit{et al.} [FENICE],
``Measurement of the total $e^+ e^- \to$~hadrons cross-section near the $e^+ e^- \to N \bar N$~threshold,''
Phys. Lett. B \textbf{365} (1996), 427-430.

\bibitem{Frabetti:2001ah}
P.~L.~Frabetti \textit{et al.} [E687],
``Evidence for a Narrow Dip Structure at 1.9~GeV/$c^{2}$ in 3 $\pi^{+} 3 \pi^{-}$ DiffractivePphotoproduction,''
Phys. Lett. B \textbf{514} (2001), 240-246.


\bibitem{Frabetti:2003pw}
P.~L.~Frabetti, H.~W.~K.~Cheung, J.~P.~Cumalat, C.~Dallapiccola, J.~F.~Ginkel, W.~E.~Johns, M.~S.~Nehring, E.~W.~Vaandering, J.~N.~Butler and S.~Cihangir, \textit{et al.}
``On the narrow dip structure at $1.9$ GeV/c$^{2}$ in diffractive photoproduction,''
Phys. Lett. B \textbf{578} (2004), 290-296.


\bibitem{Matsinos:2020cft}
E.~Matsinos,
``Determination of the masses and decay widths of the scalar-isoscalar and vector-isovector mesons below $2$ GeV,''
[arXiv:2007.13130 [hep-ph]].

\bibitem{Surovtsev:2008zza}
Y.~S.~Surovtsev and P.~Bydzovsky,
``Analysis of the pion pion scattering data and rho-like mesons,''
Nucl. Phys. A \textbf{807} (2008), 145-157.

\bibitem{Bugg:2012yt}
D.~V.~Bugg,
``Comment on \textquotedblleft{}Systematics of radial and angular-momentum Regge trajectories of light nonstrange $q\overline{q}$-states\textquotedblright{},''
Phys. Rev. D \textbf{87} (2013) no.11, 118501.


\bibitem{He:2013ttg}
L.~P.~He, X.~Wang and X.~Liu,
``Towards two-body strong decay behavior of higher $\rho$ and $\rho_3$ mesons,''
Phys. Rev. D \textbf{88} (2013) no.3, 034008.

\bibitem{Wang:2020kte}
L.~M.~Wang, J.~Z.~Wang and X.~Liu,
``Toward $e^+e^-\to \pi^+\pi^-$ annihilation inspired by higher $\rho$ mesonic states around 2.2 GeV,''
Phys. Rev. D \textbf{102} (2020) no.3, 034037.

\bibitem{Hasan:1994he}
A.~Hasan and D.~V.~Bugg,
``Amplitudes for $\bar p p \to \pi \pi$ from 0.36~GeV/c to 2.5~GeV/c,''
Phys. Lett. B \textbf{334} (1994), 215-219.



\bibitem{Ablikim:2019tqd}
M.~Ablikim \textit{et al.} [BESIII],
``Partial-wave analysis of $J/\psi \to K^+K^-\pi^0$,''
Phys. Rev. D \textbf{100} (2019) no.3, 032004.

\bibitem{Ablikim:2020wyk}
M.~Ablikim \textit{et al.} [BESIII],
``Measurement of the Born cross sections for $e^+e^- \to \eta^\prime \pi^{+}\pi^{-}$ at center-of-mass energies between $2.00$ and $3.08$\textasciitilde{}GeV,''
Phys. Rev. D \textbf{103} (2021) no.7, 072007.


\bibitem{Atkinson:1985yx}
M.~Atkinson \textit{et al.} [Omega Photon],
``Evidence for a $\omega \rho^\pm \pi^\mp$ State in Diffractive Photoproduction,''
Z. Phys. C \textbf{29} (1985), 333.

\bibitem{Bisello:1981sh}
D.~Bisello, J.~C.~Bizot, J.~Buon, A.~Cordier, B.~Delcourt and F.~Mane,
``Study of the Reaction $e^+ e^- \to 3 \pi^+ 3 \pi^-$ in the Total Energy Range 1400~MeV to 2180~MeV,''
Phys. Lett. B \textbf{107} (1981), 145-147.

\bibitem{Alde:1992wv}
A.~Alde \textit{et al.} [IHEP-IISN-LANL-LAPP-KEK],
``Study of the $\omega \pi^0$ system,''
Z. Phys. C \textbf{54} (1992), 553-558.

\bibitem{Alde:1994jm}
D.~Alde \textit{et al.} [GAMS],
``Partial wave analysis of the $\omega \pi^0$ system at high masses,''
Nuovo Cimento Soc. Ital. Fis. 107A, 1867 (1994).



\bibitem{Anisovich:1999xm}
A.~V.~Anisovich, V.~A.~Nikonov, A.~V.~Sarantsev, V.~V.~Sarantsev, C.~A.~Baker, C.~J.~Batty, D.~V.~Bugg, A.~Hasan, C.~Hodd and B.~S.~Zou, \textit{et al.}
``Analysis of $\bar p p \to \pi^- \pi^+, \pi^0 \pi^0, \eta \eta$ and $\eta \eta'$ from threshold to 2.5-GeV/c,''
Phys. Lett. B \textbf{471} (1999), 271-279.

\bibitem{Lees:2013gzt}
J.~P.~Lees \textit{et al.} [BaBar],
``Precision measurement of the $e^+e^- \to K^+ K^- (\gamma)$ cross section with the initial-state radiation method at BABAR,''
Phys. Rev. D \textbf{88} (2013) no.3, 032013.

\bibitem{Lees:2018vvb}
J.~P.~Lees \textit{et al.} [BaBar],
``Study of the process $e^+e^- \to \pi^+\pi^-\eta $ using initial state radiation,''
Phys. Rev. D \textbf{97} (2018), 052007.

\bibitem{Masjuan:2013xta}
P.~Masjuan, E.~Ruiz Arriola and W.~Broniowski,
``Reply to \textquotedblleft{}Comment on \textquoteleft{}Systematics of radial and angular-momentum Regge trajectories of light nonstrange $q\bar{q}$-states\textquoteright{} \textquotedblright{},''
Phys. Rev. D \textbf{87} (2013) no.11, 118502.


\bibitem{Chao:1992et}
K.~T.~Chao, Y.~B.~Ding and D.~H.~Qin,
``Possible phenomenological indication for the string Coulomb term and the color screening effects in the quark - anti-quark potential,''
Commun. Theor. Phys. \textbf{18} (1992), 321-326.

\bibitem{Ding:1993uy}
Y.~B.~Ding, K.~T.~Chao and D.~H.~Qin,
``Screened $Q \bar Q$ potential and spectrum of heavy quarkonium,''
Chin. Phys. Lett. \textbf{10} (1993), 460-463.

\bibitem{Wang:2018rjg}
J.~Z.~Wang, Z.~F.~Sun, X.~Liu and T.~Matsuki,
``Higher bottomonium zoo,''
Eur. Phys. J. C \textbf{78} (2018) no.11, 915.


\bibitem{Pang:2017dlw}
C.~Q.~Pang, J.~Z.~Wang, X.~Liu and T.~Matsuki,
``A systematic study of mass spectra and strong decay of strange mesons,''
Eur. Phys. J. C \textbf{77} (2017) no.12, 861.

\bibitem{Pang:2019ttv}
C.~Q.~Pang,
``Excited states of $\phi$ meson,''
Phys. Rev. D \textbf{99} (2019) no.7, 074015.



\bibitem{Pang:2018gcn}
C.~Q.~Pang, Y.~R.~Wang and C.~H.~Wang,
``Prediction for $5^{++}$ mesons,''
Phys. Rev. D \textbf{99} (2019) no.1, 014022.

\bibitem{Hao:2019fjg}
W.~Hao, G.~Y.~Wang, E.~Wang, G.~N.~Li and D.~M.~Li,
``Canonical interpretation of the $X(4140)$ state within the $^3P_0$ model,''
Eur. Phys. J. C \textbf{80} (2020) no.7, 626.

\bibitem{Zeng:1994vj}
J.~Zeng, J.~W.~Van Orden and W.~Roberts,
``Heavy mesons in a relativistic model,''
Phys. Rev. D \textbf{52} (1995), 5229-5241.


\bibitem{Close:2005se}
F.~E.~Close and E.~S.~Swanson,
``Dynamics and decay of heavy-light hadrons,''
Phys. Rev. D \textbf{72} (2005), 094004

\bibitem{Zhong:2008kd}
X.~H.~Zhong and Q.~Zhao,
``Strong decays of heavy-light mesons in a chiral quark model,''
Phys. Rev. D \textbf{78} (2008), 014029


\bibitem{Li:2008we}
D.~M.~Li and B.~Ma,
``$\eta(2225)$ observed by the BES Collaboration,''
Phys. Rev. D \textbf{77} (2008), 094021.

\bibitem{Li:2008et}
D.~M.~Li and S.~Zhou,
``Towards the assignment for the $4^1S_0$ meson nonet,''
Phys. Rev. D \textbf{78} (2008), 054013.

\bibitem{Xue:2018jvi}
S.~C.~Xue, G.~Y.~Wang, G.~N.~Li, E.~Wang and D.~M.~Li,
``The possible members of the $5^1S_0$ meson nonet,''
Eur. Phys. J. C \textbf{78} (2018) no.6, 479.

\bibitem{Wang:2017pxm}
G.~Y.~Wang, S.~C.~Xue, G.~N.~Li, E.~Wang and D.~M.~Li,
``Strong decays of the higher isovector scalar mesons,''
Phys. Rev. D \textbf{97} (2018) no.3, 034030.

\bibitem{Pan:2016bac}
T.~T.~Pan, Q.~F.~L\"u, E.~Wang and D.~M.~Li,
``Strong decays of the $X(2500)$ newly observed by the BESIII Collaboration,''
Phys. Rev. D \textbf{94} (2016) no.5, 054030.

\bibitem{Lu:2016bbk}
Q.~F.~L\"u, T.~T.~Pan, Y.~Y.~Wang, E.~Wang and D.~M.~Li,
``Excited bottom and bottom-strange mesons in the quark model,''
Phys. Rev. D \textbf{94} (2016) no.7, 074012.

\bibitem{Song:2014mha}
Q.~T.~Song, D.~Y.~Chen, X.~Liu and T.~Matsuki,
``$D_{s1}^*(2860)$ and $D_{s3}^*(2860)$ : candidates for $1D$ charmed-strange mesons,''
Eur. Phys. J. C \textbf{75} (2015) 1, 30.


\bibitem{Wang:2016krl}
J.~Z.~Wang, D.~Y.~Chen, Q.~T.~Song, X.~Liu and T.~Matsuki,
``Revealing the inner structure of the newly observed $D_2^*(3000)$,''
Phys. Rev. D \textbf{94} (2016) no.9, 094044.

\bibitem{Pang:2014laa}
C.~Q.~Pang, L.~P.~He, X.~Liu and T.~Matsuki,
``Phenomenological study of the isovector tensor meson family,''
Phys. Rev. D \textbf{90} (2014) no.1, 014001.


\bibitem{Micu:1968mk}
L.~Micu,
``Decay rates of meson resonances in a quark model,''
Nucl. Phys. B \textbf{10} (1969), 521-526.

\bibitem{LeYaouanc:1972vsx}
A.~Le Yaouanc, L.~Oliver, O.~Pene and J.~C.~Raynal,
``Naive quark pair creation model of strong interaction vertices,''
Phys. Rev. D \textbf{8} (1973), 2223-2234.


\bibitem{Blundell:1996as}
H.~G.~Blundell,
``Meson properties in the quark model: A look at some outstanding problems,''
[arXiv:hep-ph/9608473 [hep-ph]].

\bibitem{Hayne:1981zy}
C.~Hayne and N.~Isgur,
``Beyond the Wave Function at the Origin: Some Momentum Dependent Effects in the Nonrelativistic Quark Model,''
Phys. Rev. D \textbf{25}, 1944 (1982).

\bibitem{Jacob:1959at}
M.~Jacob and G.~C.~Wick,
``On the General Theory of Collisions for Particles with Spin,''
Annals Phys. \textbf{7} (1959), 404-428.








\bibitem{Dudek:2010wm}
J.~J.~Dudek, R.~G.~Edwards, M.~J.~Peardon, D.~G.~Richards and C.~E.~Thomas,
``Toward the excited meson spectrum of dynamical QCD,''
Phys. Rev. D \textbf{82} (2010), 034508


\bibitem{Johnson:2020ilc}
C.~T.~Johnson \textit{et al.} [Hadron Spectrum],
``Excited $J^{--}$ meson resonances at the SU(3) flavor point from lattice QCD,''
Phys. Rev. D \textbf{103} (2021) no.7, 074502
























\end{thebibliography}
\end{document}